\documentstyle[prl,aps,multicol,epsf,epsfig,psfig]{revtex}
\begin{document}
\title{Uncertainty is complementary to Complementarity}
\author{Ole Steuernagel,
\\
Department of Physical Sciences, University of Hertfordshire,
Hatfield, Herts, AL10 9AB, UK
\\}
\date{\today}
\twocolumn{
\maketitle
\begin{abstract}
For any ideal two-path interferometer it is shown that the
wave-particle duality of quantum mechanics implies Heisenberg's
uncertainty relation and vice versa. It is conjectured that
complementarity and uncertainty are two aspects of the same
general principle.
\end{abstract}
\narrowtext
Bohr's principle of complementarity, applied to a two-path
interferometer, describes the nature of a quantum system as being
'dualistic' in its particle and wave aspects~\cite{Wheeler.buch}.
Although Bohr originally intended the principle of complementarity
to be of greater generality than the wave-particle duality of
quantum mechanics both concepts are now often treated as
equivalent~\cite{Wheeler.buch,Feynman.buch}. In the famous debates
between Einstein and Bohr in the late 1920's~\cite{Wheeler.buch}
complementarity was contested but finally the argument has settled
in its favor~\cite{Wheeler.buch,Feynman.buch}. The foundation of
the concept of complementarity, however, is still argued
upon~\cite{Scully91.nature,Zajonc91,Englert95.nature,Englert96,Knight98.nature,Duerr98,NewScientist98,Stern90,Bhandari92,Tan93,Storey94,Storey95.nature,Wiseman95.nature}.
Whereas Bohr invariably used different versions of Heisenberg's
uncertainty relation in refuting Einstein's attempts to disprove
the concept of complementarity~\cite{Wheeler.buch}, a theoretical
proposal from 1991~\cite{Scully91.nature} and an experiment
performed last year~\cite{Duerr98} have been interpreted as
showing that complementarity can arise without
uncertainty~\cite{Knight98.nature,Duerr98,NewScientist98}. The
question therefore remains as to whether complementarity always
arises from Heisenberg's uncertainty
principle~\cite{Bhandari92,Tan93,Storey94,Storey95.nature,Wiseman95.nature}
or whether a 'more fundamental mechanism' -- entanglement without
uncertainty -- can be at
work~\cite{Englert96,Knight98.nature,NewScientist98}?
\\
To answer this question, optimal two-path interferometers are
analyzed; this is all that is needed since we are only interested
in fundamental physical limits, thus neglecting unbalanced, lossy,
and otherwise imperfect setups. Moreover, two paths, described by
two quantum mechanical modes, capture all the essential physics,
including all double slit setups and all systems currently under
discussion~\cite{Scully91.nature,Zajonc91,Englert95.nature,Englert96,Knight98.nature,Duerr98,NewScientist98,Stern90,Bhandari92,Tan93,Storey94,Storey95.nature,Wiseman95.nature}.
\\
In a two-mode interferometer two 'paths' does not necessarily
refer to spatially separated paths, it can, for instance, mean two
different spin states in Ramsey
interferometry~\cite{Scully91.nature,Duerr98}. Heisenberg's
position-momentum uncertainty should therefore not always be
appropriate for the description of the loss of interference
fringes when determining the path of a quantum
particle~\cite{Scully91.nature,Duerr98,Bhandari92}. This has
recently been
debated~\cite{Scully91.nature,Englert95.nature,Bhandari92,Tan93,Storey94,Storey95.nature,Wiseman95.nature}
but seems to be resolved now~\cite{Duerr98}: complementarity can
be enforced without invoking {\em position-momentum uncertainty}.
\\
We will, however, see that the assumption that path and wave
measurements ($\hat P$ and $\hat W$) are complementary always
leads to the {\em general Heisenberg--Robertson uncertainty
relation}~\cite{Wheeler.buch,Robertson29,Maassen88}
\begin{eqnarray}
\Delta \hat P \cdot \Delta \hat W \geq \frac{1}{2} \; |\langle
[\hat P,\hat W] \rangle |
 \; ,
\label{heisrob.unc}
\end{eqnarray}
and that the ensuing uncertainties are sufficient to make
inaccessible either the path or the wave aspect of the quantum
system: complementarity implies uncertainty and this uncertainty
enforces complementarity.
\\
In an optimal interferometer the two paths are identified with two
quantum mechanical modes and can, without loss of generality, be
assigned the two basis states of a formal spin-1/2 system
$|\psi_+\rangle =(1,0)$ and $|\psi_-\rangle = (0,1)$.  To scan the
interference pattern the relative phase $\phi$ between these two
basis states is changed by a phase shifter $\hat \Phi \;( =
\exp[-i \; \hat \sigma_z \; \phi/2]$ without loss of
generality~\cite{Englert96}). When measuring the particle aspect,
i.e. experimentally discriminating the paths, see Fig.1., the
paths are assigned two different values~$p_\pm$. Using the above
representation for $|\psi_\pm\rangle$ we hence find that the path
operator $\hat P $ has the general form $\hat P = p_+
|\psi_+\rangle\langle \psi_+ | + p_- |\psi_- \rangle \langle
\psi_- | $ which, after suitable rescaling and without loss of
generality, leads to our choice $p_\pm =\pm 1$ and the customary
form~\cite{Englert96}
\begin{eqnarray}
\hat P =\left(
 \begin{array}{cc}  1 & 0 \\  0 & -1  \end{array}
 \right) = \hat \sigma_z
 \, ,
\label{poper.unc}
\end{eqnarray}
where the $\sigma$'s are the Pauli-matrices for spin-1/2.
Analogous considerations hold for the operator~$\hat W$ describing
the measurement of the wave features of the quantum particle. This
is easiest seen from the fact that the determination of the
interference pattern also amounts to a path measurement,~$\hat
\sigma_z^{after} $, after the final beam-splitter~$\hat B$, see
Fig.1. We conclude that, without loss of generality, $\hat W$ has
two eigenvalues $w_\pm = \pm 1$ with associated eigenvectors $|
\omega_\pm \rangle $.
\\
What is the general form of $\hat W$ in terms of the basis
vectors~$|\psi_\pm \rangle $ of $\hat P$? To find this connection
between path and wave measurement all we need to use is the fact
that they are complementary to each other:
\\
"We say that two variables are 'complementary' if precise
knowledge of one of them implies that all possible outcomes of
measuring the other one are equally
probable."~\cite{Scully91.nature}
\\
With our convention~$p_\pm =\pm 1$ the expectation value of a
path-measurement~$\hat P$, after a preceding measurement has
projected the particle into a wave-eigenstate~$| \omega_\pm
\rangle $, must therefore be zero
\begin{eqnarray}
\langle \omega_+ | \hat P |\omega_+ \rangle = 0 = \langle \omega_-
| \hat P |\omega_- \rangle
 \, . \label{pzero.unc}
\end{eqnarray}
The complementary statement is
\begin{eqnarray}
\langle \psi_+ | \hat W |\psi_+ \rangle = 0 = \langle \psi_- |
\hat W |\psi_- \rangle
 \, . \label{wzero.unc}
\end{eqnarray}
Both statements are in accord with other quantifications of
complementarity suggested in recent
years~\cite{Wooters79,Mandel91.opt,Englert96}. Using the
orthonormality condition $\langle \omega_+ | \omega_- \rangle = 0$
we can solve equation~(\ref{pzero.unc}) and find
\begin{eqnarray}
|\omega_\pm \rangle = \frac{1}{\sqrt{2}} \; (\pm
\exp[-i\phi_0/2],\exp[i\phi_0/2])
 \, .
\label{balstat.unc}
\end{eqnarray}
The phase angle~$\phi_0$ depends on the details of the
interferometric setup. Note that the $|\omega_\pm \rangle$ are
balanced, i.e., the particle is equally probably found in the
upper or lower path of the interferometer. Only such states allow
for maximum contrast of the interference
pattern~\cite{Mandel91.opt}. Consequently all states $ |\phi
\rangle $ we encounter in an optimal interferometer need to have
this balanced form
\begin{eqnarray}
|\phi \rangle = \frac{1}{\sqrt{2}} \;
(\exp[-i\phi/2],\exp[i\phi/2])
 \, .
\label{bala.unc}
\end{eqnarray}
This also follows from the fact that a pure phase shift
operation~$\hat \Phi $ applied to $| \omega_\pm \rangle$ can only
generate states of the form~(\ref{bala.unc}).
Using~(\ref{balstat.unc}) we find
\begin{eqnarray}
\hat W = w_+ |\omega_+\rangle \langle \omega_+ | + w_- |\omega_-
\rangle \langle \omega_- |
\\
= \cos \phi_0 \; \hat \sigma_x + \sin \phi_0 \; \hat \sigma_y
 \, .
\label{woper.unc}
\end{eqnarray}
This form of $\hat W $ might look unfamiliar but it describes,
what we would expect, a path measurement after the last beam
splitter $\hat W = \hat B^\dagger \hat \sigma_z \hat B = \hat
\sigma_z^{after} $ expressed in terms of the basis inside the
interferometer~\cite{vertausch}. Using~(\ref{bala.unc})
and~(\ref{woper.unc}) we find the expectation value of~$ \hat W $
is
\begin{eqnarray}
\langle \phi | \hat W | \phi \rangle = \cos (\phi-\phi_0)
 \, ,
\label{wval.unc}
\end{eqnarray}
the expected classical interference pattern. This confirms that
the above form of $\hat W$ indeed represents the sought-after
'wave-operator'. We have succeeded in deriving $\hat W$ from
equation~(\ref{pzero.unc}) -- the principle of complementarity --
alone.
\\
Obviously $\hat P$ and $\hat W$ do not commute $([\hat
\sigma_z,\hat \sigma_x]=2i\hat \sigma_y$ and $[\hat \sigma_z,\hat
\sigma_y]=-2i\hat \sigma_x)$, hence, the uncertainty
relation~(\ref{heisrob.unc}) becomes

\begin{eqnarray}
\Delta \hat P \cdot \Delta \hat W \geq | \langle \cos \phi_0 \;
\hat \sigma_y - \sin \phi_0 \; \hat \sigma_x \rangle |
 \, .
\label{heise.unc}
\end{eqnarray}
For an optimal interferometer, i.e. for a balanced
state~(\ref{bala.unc}), this relation assumes the specific form
\begin{eqnarray}
\Delta_\phi \hat P \cdot \Delta_\phi \hat W \geq | \sin( \phi
-\phi_0) |
 \, .
\label{heisespec.unc}
\end{eqnarray}
In separate calculations one can easily confirm that $\Delta_\phi
\hat P = 1$ and $\Delta_\phi \hat W=| \sin(\phi - \phi_0 )|$, a
pictorial representation of these uncertainties is given in Fig.2.
One might wonder whether Eq.~(\ref{heisespec.unc}) describes a
valid Heisenberg uncertainty relationship since it gives a
vanishing lower bound for states for which $\phi - \phi_0 $ is an
integer multiple of $\pi$. But this is no reason for worry, it
reflects the well-known fact that this bound vanishes for
eigenstates $(|\phi \rangle = | \omega_\pm \rangle $) of the
considered observables~\cite{Maassen88}. For the important case of
greatest interferometric sensitivity, when $|\delta \langle \hat W
\rangle /\delta \phi | = |\delta \cos ( \phi -\phi_0)/\delta \phi
| $ is maximal, the uncertainty is maximal as well.
\\
For any two-path interferometer we have derived the uncertainty
relation that goes together with complementarity. Fig.2. shows
that this uncertainty relation quantifies 'just what is needed':
the variances have the magnitudes required to project the quantum
state $|\phi \rangle$ into the eigenstates of the respective
measurement. This projection uncertainty completely destroys the
complementary information; when measuring the interference pattern
all path knowledge is lost~(\ref{pzero.unc}) and vice
versa~(\ref{wzero.unc}): uncertainty enforces complementarity.
\\
Uncertainty relations~(\ref{heise.unc}) and ~(\ref{heisespec.unc})
hold for any two-mode duality experiment. For the sake of
generality we have not specified the coupling that entangles
particle and detector, but, once the coupling is specified,
relation~(\ref{heise.unc}) encourages us to search for other
problem-specific uncertainty relations that allow us to get a
deeper understanding of the respective
mechanism~\cite{Stern90,Luis98} and even to discover new
physics~\cite{Bhandari92}.
\\
Our analysis also sheds a fresh light on recent discussions: in
particular we find that the claim that complementarity could be
enforced without the unavoidable 'measurement disturbances'
characteristic of the principle of
uncertainty~\cite{Scully91.nature,Knight98.nature,NewScientist98}
is not substantiated. Also the related idea that complementarity
is a 'deeper principle' than
uncertainty~\cite{Englert96,Knight98.nature,NewScientist98} is in
disagreement with our present findings.
\\
The analysis of complementarity and uncertainty presented here
shows that for any two-path interferometer uncertainty and
complementarity mutually imply each other; this leads me to the
general conjecture that uncertainty and complementarity are two
aspects of the same principle: Bohr's principle of complementarity
is complementary to Heisenberg's uncertainty principle.
%
%

%
\acknowledgments I wish to thank John Vaccaro, Howard Wiseman, Jim
Collett and Koji Murakawa for lively discussions.
\newpage
\begin{figure}[htb]
%
\epsfverbosetrue \epsfxsize=2in \epsfysize=2in
\epsffile[-130 -50 420 500]{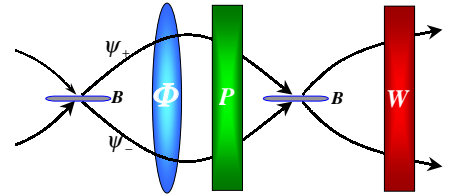}
\caption{A cartoon of a general two-mode interferometer: to create
a balanced state~(\protect\ref{bala.unc}) inside the
interferometer the particle is, say, entering through the lower
port of the first semi-transparent beam splitter $B$. The
subsequent phase-shifter $\Phi$ allows us to scan the interference
pattern which is measured by $\hat W$, this is usually done after
the second beam-splitter as a path measurement, $\hat
\sigma_z^{after}$, but could be done inside the
interferometer~\protect\cite{vertausch}. Alternatively, the path
of the particle can be determined, most conveniently by measuring
$\hat P = \hat \sigma_z$ inside the interferometer. }
%
\end{figure}
%
\vspace{2.5cm}
\begin{figure}[htb]
%
\epsfverbosetrue \epsfxsize=2.5in \epsfysize=1.5in

\epsffile[-30 150 420 430]{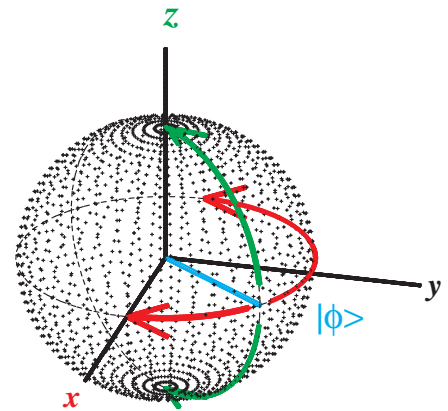}
\caption{A Bloch-sphere representation of the general balanced
state $| \phi \rangle $ of Eq.~(\ref{bala.unc}) in an optimal
interferometer (we assume $\phi_0 =0$ without loss of generality).
Since $| \phi \rangle $ is balanced it lies in the equatorial
$xy$-plane. This is how the measurement uncertainties enforce
complementarity:
\\
If the path measurement~$\hat P$ is performed first the state $|
\phi \rangle $ is projected out of the $xy$-plane onto the
$z$-axis (green arrows) and equally often ends up pointing 'North'
or 'South'. The ensuing variance of this process is unity.
\\
If the interference measurement~$\hat W$ is performed first the
state is projected onto the $x$-axis (red arrows), the variance
then depends on the angle and equals~$|\sin\phi|$.
\\
In either case one measurement completely randomizes the outcome
of the other, the variables are hence complementary.
}
\end{figure}

\begin{thebibliography}{99}


\bibitem{Wheeler.buch} J. A. Wheeler and W. H. Zurek, {\em Quantum Theory and
Measurement}, (Princeton University Press, 1983).

\bibitem{Feynman.buch} R. P. Feynman, R. B. Leighton and M. Sands,
{\em The Feynman Lectures on Physics}, (Addison-Wesley, Reading,
1965).

\bibitem{Scully91.nature} M. O. Scully, B.-G. Englert, and H. Walther,
{\em Quantum optical tests of complementarity}, Nature {\bf 351},
111 (1991).

\bibitem{Zajonc91} A. G. Zajonc, L. J. Wang, X. Y. Zou, and L. Mandel,
{\em Quantum eraser}, Nature {\bf 353}, 507 (1991).

\bibitem{Englert95.nature} B.-G. Englert, M. O. Scully, and H. Walther,
{\em Complementarity and Uncertainty}, Nature {\bf 375}, 367
(1995).

\bibitem{Englert96} B.-G. Englert,
{\em Fringe Visibility and Which-Way Information: An Inequality},
Phys. Rev. Lett. {\bf 77}, 2154 (1996).

\bibitem{Knight98.nature} P. L. Knight, {\em Where the weirdness comes from}, Nature {\bf
395}, 12 (1998).

\bibitem{Duerr98} S. D\"urr, T. Nonn and G. Rempe,
{\em Origin of quantum-mechanical complementarity probed by a
'which-way' experiment in an atom interferometer}, Nature {\bf
395}, 33 (1998).

\bibitem{NewScientist98} M. Buchanan,
{\em An end to uncertainty}, New Scientist, 25 (6 March 1999).

\bibitem{Stern90} A. Stern, Y. Aharonov, and Y. Imry, {\em Phase uncertainty and
loss of interference: A general picture}, Phys. Rev. A {\bf 41},
3436 (1990).

\bibitem{Bhandari92} R. Bhandari,
{\em Decoherence due to the Geometric Phase in a "Welcher-Weg"
Experiment}, Phys. Rev. Lett. {\bf 69}, 3720 (1992).

\bibitem{Tan93} S. M. Tan and D. F. Walls, {\em Loss of coherence in interferometry},
Phys. Rev. A {\bf 47}, 4663 (1993).

\bibitem{Storey94} E. P. Storey, S. M. Tan, M. J. Collet, and D. F. Walls, {\em
Path detection and the uncertainty principle}, Nature {\bf 367},
626 (1994).

\bibitem{Storey95.nature} E. P. Storey, S. M. Tan, M. J. Collet, and D. F. Walls, {\em
Complementarity and Uncertainty: Reply}, Nature {\bf 375}, 368
(1995).

\bibitem{Wiseman95.nature} H. Wiseman and F. Harrison,
{\em Uncertainty over Complementarity?}, Nature {\bf 377}, 584
(1995).

\bibitem{Robertson29} H. P. Robertson, {\em The uncertainty
principle}, Phys. Rev. {\bf 34}, 163 (1929).

\bibitem{Maassen88} H. Maassen and J. B. M. Uffink,
{\em Generalized Entropic Uncertainty Relations}, Phys. Rev. Lett.
{\bf 60}, 1103 (1988).

\bibitem{Wooters79} W. K. Wootters and W. H. Zurek, {\em Complementarity
in the double-slit experiment: Quantum nonseparability and a
quantitative measure of Bohr's principle}, Phys. Rev. D {\bf 19},
473 (1979).

\bibitem{Mandel91.opt} L. Mandel,
{\em Coherence and indistinguishability}, Opt. Lett. {\bf 16},
1882 (1991).

\bibitem{vertausch} In principle the interference pattern measurement can be
performed measuring $ \cos \phi_0 \; \hat \sigma_x + \sin \phi_0
\; \hat \sigma_y $ inside the interferometer rather than $ \hat
\sigma_z^{after} $ after the second beam splitter, an analogous
statement applies to $\hat P$~\cite{Englert96}.

\bibitem{Luis98} A. Luis and L. L. S\'anchez-Soto,
{\em Complementarity Enforced by Random Classical Phase Kicks},
Phys. Rev. Lett. {\bf 81}, 4031 (1998).

\end{thebibliography}
\end{document}